\input amstex
\magnification 1200
\TagsOnRight
\def\qed{\ifhmode\unskip\nobreak\fi\ifmmode\ifinner\else
 \hskip5pt\fi\fi\hbox{\hskip5pt\vrule width4pt
 height6pt depth1.5pt\hskip1pt}}
 \def\adots{\mathinner{\mkern2mu\raise1pt\hbox{.}
\mkern3mu\raise4pt\hbox{.}\mkern1mu\raise7pt\hbox{.}}}
\def\sdots{\mathinner{
     \mskip.01mu\raise1pt\vbox{\kern1pt\hbox{.}}
     \mskip.01mu\raise3pt\hbox{.}
     \mskip.01mu\raise5pt\hbox{.}
     \mskip1mu}}
\NoBlackBoxes
\baselineskip 20 pt
\parskip 5 pt

\centerline {\bf EXACT SOLUTIONS TO}
\centerline {\bf THE SINE-GORDON EQUATION}

\vskip 10 pt
\centerline {Tuncay Aktosun}
\vskip -6 pt
\centerline {Department of Mathematics}
\vskip -6 pt
\centerline {University of Texas at Arlington}
\vskip -6 pt
\centerline {Arlington, TX 76019-0408, USA}

\centerline{Francesco Demontis and Cornelis van der Mee}
\vskip -6 pt
\centerline{Dipartimento di Matematica e Informatica}
\vskip -6 pt
\centerline{Universit\`a di Cagliari}
\vskip -6 pt
\centerline{Viale Merello 92}
\vskip -6 pt
\centerline{09123 Cagliari, Italy}

\vskip 10 pt

\noindent {\bf Abstract}:
A systematic method is presented to provide various equivalent solution formulas
for exact solutions to the sine-Gordon equation. Such solutions are analytic in the
spatial variable $x$ and the temporal variable
$t,$ and they are exponentially asymptotic to integer multiples
of $2\pi$ as $x\to\pm\infty.$ The solution formulas are expressed explicitly in terms of
a real triplet of constant matrices. The method presented is generalizable to other
integrable evolution equations where the inverse scattering transform is applied
via the use of a Marchenko integral equation. By expressing the kernel
of that Marchenko equation as a matrix exponential in terms of the matrix triplet
and by exploiting the separability of that
kernel, an exact solution formula to the Marchenko equation is derived, yielding
various equivalent exact solution formulas for the sine-Gordon equation.


\newpage

\noindent {\bf 1. INTRODUCTION}
\vskip 3 pt

Our goal in this paper is to derive, in terms of a triplet
of constant matrices, explicit
formulas for exact solutions to the
sine-Gordon equation
$$u_{xt}=\sin u,\tag 1.1$$
where $u$ is real valued and the subscripts denote
the partial derivatives with respect to
the spatial coordinate $x$ and the temporal
coordinate $t.$
Under the transformation
$$x\mapsto ax+\displaystyle\frac{t}{a},\qquad t\mapsto ax-\displaystyle\frac{t}{a},$$
where $a$ is a positive constant, (1.1) is transformed into the
alternate form
$$u_{xx}-u_{tt}=\sin u,\tag 1.2$$
and hence our explicit formulas can easily be modified to obtain
explicit solutions to (1.2) as well. Let us note that
one can omit a multiple of $2\pi$ from any solution to
(1.1). We are interested in solutions to (1.1) where $u_x(x,t)\to 0$
as $x\to\pm\infty$ for each fixed $t,$ and hence without any loss of generality
we will normalize our solutions so that $u(x,t)\to 0$
as $x\to +\infty.$

The sine-Gordon equation arises
in applications as diverse as the description of surfaces of constant mean
curvature [10,16], one-dimensional crystal dislocation theory
[17,23,40,41], magnetic flux propagation in Josephson junctions (gaps between
two superconductors)
[29,31], condensation of charge density
waves [11,22,36], wave propagation in
ferromagnetic materials [19,27,30], excitation
of phonon modes [35], and propagation of deformations along the DNA double
helix [18,26,38,43].

The literature on exact solutions
to (1.1) is large, and we will
mention only a few and refer the reader to those
references and further ones therein.
For a positive constant $a,$ by substituting
$$u(x,t)=4\,\tan^{-1}\left(\dfrac{U(ax+a^{-1}t)}{V(ax-a^{-1}t)}\right),
\tag 1.3$$
into (1.1) and solving the resulting partial
differential equations for $U$ and $V$, Steuerwald
[42] has catalogued many exact solutions to the
sine-Gordon equation in terms of elliptic functions. Some
of these solutions, including the one-soliton solution, two-soliton solutions
modeling a soliton-soliton and soliton-antisoliton collision, and the breather
solution, can be written in terms of elementary functions [25,37], while
the $n$-soliton solutions can be expressed as in (1.3) where $U$ and $V$
are certain determinants [34,39,45]. The same separation
technique can also be used to find exact solutions
to the sine-Gordon equation on finite $(x+t)$-intervals
[13].
Solutions to the sine-Gordon equation with initial data specified on invariant
algebraic manifolds of conserved quantities can be written explicitly in terms
of Jacobi theta functions [24]. The ordered exponential integrals
appearing in such solutions can be evaluated explicitly [9,28].
Let us also mention that some exact solutions to the
sine-Gordon equations can be obtained via the Darboux or
B\"acklund transformations [21,37] from already known exact solutions.

The sine-Gordon equation was the fourth nonlinear partial
differential equation whose initial-value
problem was discovered [2,3] to be solvable
by the inverse scattering transform method.
This method associates (1.1) with the
first-order system of ordinary differential equations
$$\cases \aligned\displaystyle\frac{d\xi}{dx} &=-i\lambda \xi-\displaystyle\frac{1}{2}\,u_x(x,t)\,\eta,\\
\displaystyle\frac{d\eta}{dx} &=\displaystyle\frac{1}{2}\,u_x(x,t)\,\xi+i\lambda \eta,\endaligned\endcases
\tag 1.4$$
where $u_x$ appears in the coefficients as a potential.
By exploiting the one-to-one correspondence between
$u_x$ and the corresponding scattering data for (1.4),
the inverse scattering transform method determines
the time evolution
$u(x,0)\mapsto u(x,t)$ for (1.1)
with the help of the solutions to the direct and inverse scattering
problems for (1.4). The direct scattering problem for (1.4)
amounts to finding the scattering coefficients (related to
the asymptotics of scattering solutions to
(1.4) as $x\to\pm\infty$) when $u(x,t)$ is known for all $x.$
On the other hand, the inverse scattering problem consists of finding
$u(x,t)$ from an appropriate set of scattering data for (1.4).

In this paper we provide several, but equivalent,
explicit formulas for exact solutions to (1.1).
The key idea to obtain
such explicit formulas is to express the kernel of a related
Marchenko integral equation arising
in the inverse scattering problem for (1.4) in terms of a
real triplet $(A,B,C)$ of constant matrices and by using
matrix exponentials.
Such explicit formulas provide a compact and concise way to express
our exact solutions, which can equivalently
be expressed in terms
of exponential, trigonometric (sine and cosine),
and polynomial functions of $x$ and $t.$ This can be done by ``unpacking" matrix
exponentials in our explicit formulas. As the matrix size
increases, the unpacked expressions become very long.
However, such expressions can be evaluated explicitly
for any matrix size either by hand or by using a symbolic
software package such as Mathematica. One of the
powerful features of our method comes from the fact that our concise
and compact explicit solution formulas are valid for any matrix
size in the matrix exponentials involved. In some other
available methods, exact solutions are attempted
in terms of elementary functions without the use of
matrix exponentials, and hence exact solutions
produced by such other methods will be relatively simple and
we cannot expect those methods to produce our solutions
when the matrix size is large.

Our method is generalizable and applicable to obtain similar explicit
formulas for exact solutions to other integrable
nonlinear partial differential equations, where a Marchenko integral equation
is used to solve a related inverse scattering problem.
We refer the reader to [5-7,14,15], where similar ideas are
used to obtain explicit formulas for exact solutions
to the Korteweg-de Vries equation on the half line
and to the focusing nonlinear Schr\"odinger equation and
its matrix generalizations.

In our method,
with the help
of the matrix triplet and matrix exponentials, we easily establish
the separability of the kernel of the relevant
Marchenko integral equation and thus solve it exactly by using
linear algebra. We then obtain our exact solutions to the
sine-Gordon equation by a simple integration
of the solution to the Marchenko equation. Our method easily
handles complications arising from the presence of
non-simple poles of the transmission coefficient in the related
linear system (1.4). Dealing with non-simple poles without the use
of matrix exponentials is very complicated, and this issue
has also been a problem [33,44] in solving other integrable nonlinear
partial differential equations such as the nonlinear Schr\"odinger
equation.

Our paper is organized as follows. In Section 2 we
establish our notation, introduce the relevant Marchenko integral equation, and mention
how a solution to the sine-Gordon equation is obtained from the solution to
the Marchenko equation by using the inverse scattering
transform method. In Section 3 we outline the solution to
the Marchenko integral equation when its kernel is represented
in terms of a triplet of matrices $(A,B,C)$ and thus we derive
two solution formulas for exact solutions to
the sine-Gordon equation.
In Sections~4 and 5 we show that our explicit solution formulas hold when
the input matrix triplets come from a larger family;
we show that our solution formulas
in the more general case can be
obtained by constructing two auxiliary constant matrices $Q$ and $N$
satisfying the respective Lyapunov equations given in Section 4,
or equivalently by constructing an auxiliary constant matrix $P$
satisfying the Sylvester equation given in Section 5.
In Section 4 we also show that
the matrix triplet $(A,B,C)$ used as input to construct
our exact solutions to the sine-Gordon equation can
be chosen in various equivalent ways and
we prove that
our exact solutions
are analytic on the $xt$-plane.
In Section 5 we also
explore the relationship between the Lyapunov equations and the
Sylvester equation and show how their solutions are related to each
other in a simple but interesting way. In that section we also show that
the two solution formulas derived in Section 3 are equivalent.
In Section~6 we show
that those two equivalent solution
formulas
can be represented in other equivalent forms.
In Section~7 we evaluate the square of the
spatial derivative of our solutions to (1.1)
by providing some explicit formulas in terms of the
matrix triplet $(A,B,C),$ and we evaluate
the asymptotics of our exact solutions
as $x\to-\infty$ for each fixed $t.$
In Section~8 we show that the reflection coefficients associated
with such solutions are zero, and we also evaluate explicitly
the corresponding transmission coefficient.
Finally, in Section 9 we provide some specific examples of
our exact solutions and their snapshots.

Let us remark on the logarithm and inverse tangent functions we
use throughout our paper. The log function we use is the principal
branch of the complex-valued logarithm function and it has its branch cut
along the negative real axis while $\log(1)=0.$ The $\tan^{-1}$ function we use is the single-valued
branch related to the principal branch of the logarithm as
$$\tan^{-1} z=\displaystyle\frac{1}{2i}\log\left(
\displaystyle\frac{1+iz}{1-iz}\right),\qquad \log z=2i\tan^{-1}\left(
\displaystyle\frac{i(1-z)}{1+z}\right),\tag 1.5$$
and its branch cut is $(-i\infty,-i]\cup[i,+i\infty).$
For any square matrix $M$ not having eigenvalues
on that branch cut, we define
$$\tan^{-1}M:=\displaystyle\frac{1}{2\pi i}\int_\gamma
dz\, [\tan^{-1} z](z I-M)^{-1},\tag 1.6$$
where the contour $\gamma$ encircles each eigenvalue of $M$ exactly
once in the positive direction
and avoids the branch cut of $\tan^{-1}z.$
If all eigenvalues of $M$ have modulus less than $1,$
we then have the familiar series expansion
$$\tan^{-1}M:=M-\displaystyle\frac{1}{3}M^3+
\displaystyle\frac15 M^5-\displaystyle\frac17 M^7+\dots.$$
For real-valued $h(x)$ that vanishes as $x\to+\infty,$
the function
$\tan^{-1}(h(x))$ always has range $(-\pi/2,\pi/2)$
when $x$ values are restricted to $(x_0,+\infty)$
for some large $x_0$ value; our $\tan^{-1}(h(x))$ is
the continuous extension of that piece from
$x\in(x_0,+\infty)$ to $x\in(-\infty,+\infty).$

\vskip 10 pt
\noindent {\bf 2. PRELIMINARIES}
\vskip 3 pt

In this section we briefly review the scattering and inverse scattering
theory for (1.4) by introducing the scattering coefficients
and a Marchenko integral equation associated with (1.4). We assume that
$u$ is real valued and that $u_x$ is integrable in $x$ for each fixed $t.$
We also mention how a solution to the sine-Gordon equation is
obtained from the solution to the Marchenko equation.
We refer the reader to the
generic references such as [1,4,25,32] for the details.

Two linearly independent
solutions to (1.4) known as the Jost solutions from the left and from the
right, denoted by $\psi(\lambda,x,t)$ and $\phi(\lambda,x,t),$ respectively, are
those solutions satisfying the respective spatial asymptotics
$$\psi(\lambda,x,t)=\bmatrix 0\\
\noalign{\medskip}
e^{i\lambda x}\endbmatrix+o(1),\qquad x\to+\infty,\tag 2.1$$
$$\phi(\lambda,x,t)=\bmatrix e^{-i\lambda x}\\
\noalign{\medskip}
0
\endbmatrix+o(1),\qquad x\to-\infty.$$
The scattering coefficients for (1.4), i.e. the transmission coefficient
$T,$ the right reflection coefficient $R,$ and the left
reflection coefficient $L,$ can be defined through the spatial asymptotics
$$\psi(\lambda,x,t)=\bmatrix \displaystyle\frac{L(\lambda,t)\,e^{-i\lambda x}}{T(\lambda)}\\
\noalign{\medskip}
\displaystyle\frac{e^{i\lambda x}}{T(\lambda)}\endbmatrix+o(1),\qquad x\to-\infty,\tag 2.2$$
$$\phi(\lambda,x,t)=\bmatrix \displaystyle\frac{e^{-i\lambda x}}{T(\lambda)}\\
\noalign{\medskip}
\displaystyle\frac{R(\lambda,t)\,e^{i\lambda x}}{T(\lambda)}
\endbmatrix+o(1),\qquad x\to+\infty,$$
where
$T$ does not depend on $t,$ and $R$ and $L$ depend on $t$ as
$$R(\lambda,t)=R(\lambda,0)\,e^{-it/(2\lambda)},\qquad
L(\lambda,t)=L(\lambda,0)\,e^{it/(2\lambda)}.$$
We recall that a bound state
corresponds to a square-integrable solution to (1.4) and such solutions
can only occur at the poles of the meromorphic extension of $T$
to the upper half complex $\lambda$-plane denoted by $\bold C^+.$
Because $u(x,t)$ is real valued, such poles can occur either on the positive
imaginary axis, or for each pole not on the positive imaginary axis there
corresponds a pole symmetrically located with respect
to the imaginary axis. Furthermore, such
poles are not necessarily simple.
If $u_x$ is integrable in $x$ for each fixed $t$ and if
the transmission coefficient $T$ is continuous for
real values of $\lambda,$ it can be proved by elementary means that
the number of such poles and their multiplicities are finite.

With the convention $u(x,t)\to 0$ as $x\to+\infty,$
it is known that $u(x,t)$ in (1.4) can be determined as
$$u(x,t)=-4\int_x^\infty dr\,K(r,r,t),\tag 2.3$$
or equivalently we have
$$u_x(x,t)=4K(x,x,t),$$
where $K(x,y,t)$ is the solution to the Marchenko integral equation
$$K(x,y,t)-\Omega(x+y,t)^*+\int_x^\infty dv\int_x^\infty dr \,
K(x,v,t)\,\Omega(v+r,t)\,\Omega(r+y,t)^*=0,\qquad y>x,\tag 2.4$$
where the asterisk is used to denote complex conjugation
(without taking the matrix transpose) and
$$\Omega(y,t)=\displaystyle\frac{1}{2\pi}\int_{-\infty}^\infty d\lambda
\,R(\lambda,t)\,e^{i\lambda y}+\displaystyle\sum_{j=1}^n
c_j\,e^{i\lambda_j y-it/(2\lambda_j)},\tag 2.5$$
provided the poles $\lambda_j$ of the transmission coefficient
are all simple.

\noindent The inverse scattering transform procedure can be summarized
via the following diagram:

\vskip 10 pt
\qquad\qquad {$\CD u(x,0) @> {} >> u_x(x,0) @>\text{direct scattering at } t=0>> \{R(\lambda,0),
\{\lambda_j,c_j\}\} \\
@V{\text{sine-Gordon solution}}VV     @.  @VV{   \text{time evolution}  }V
 \\
u(x,t) @<< {} < u_x(x,t)
@<<\text{inverse scattering at } t< \{R(\lambda,t),
\{\lambda_j,c_j\,e^{-it/(2\lambda_j)}\}\}
\endCD$}
\vskip 10 pt

We note that in general the summation term in (2.5) is much more complicated,
and the expression we have provided for it in (2.5) is valid only when
the transmission coefficient $T$ has simple poles at $\lambda_j$
with $j=1,\dots,n$
on $\bold C^+.$
In case of bound states with nonsimple poles,
it is unknown to us if the norming constants with the appropriate
time dependence have ever been presented in the literature.
Extending our previous results for the nonlinear Schr\"odinger equation
[6,7,12,14] to the sine-Gordon equation, it is possible to obtain the
norming constants with appropriate dependence on the parameter $t$
in the most general case, whether the bound-state
poles occur on the positive imaginary axis or occur
pairwise located symmetrically with respect to the positive
imaginary axis, and whether any such poles are simple or have multiplicities.
In fact, in Section~8 we present the norming constants and their proper
time dependence on $t$ as well as the most general form of the
summation term that should appear in (2.5).

When $u$ is real valued, it is known that for real $\lambda$ we have
$$R(-\lambda,t)=R(\lambda,t)^*,\qquad
L(-\lambda,t)=L(\lambda,t)^*,\qquad
T(-\lambda)=T(\lambda)^*.$$
Because $u$ is real valued, as we verify in Section~3, both the kernel
$\Omega(y,t)$ and the solution $K(x,y,t)$ in (2.4) are also
real valued, i.e.
$$\Omega(y,t)^*=\Omega(y,t),\tag 2.6$$
$$K(x,y,t)^*=K(x,y,t).\tag 2.7$$

\vskip 10 pt
\noindent {\bf 3. EXPLICIT SOLUTIONS TO THE SINE-GORDON EQUATION}
\vskip 3 pt

Our goal in this section is to obtain some
exact solutions to the sine-Gordon equation in terms of a triplet of
constant matrices. Following the main idea of [6,7] we will replace
the summation term in (2.5) by a compact expression in terms of
a matrix triplet $(A,B,C),$ i.e. we will replace $\Omega(y,t)$ when $R=0$ by
$$\Omega(y,t)=Ce^{-Ay-A^{-1}t/2}B,\tag 3.1$$
where $A,$ $B,$ $C$ are real and constant matrices
of sizes $p\times p,$ $p\times 1,$ and $1\times p,$ respectively,
for some positive integer $p.$

Recall that any rational function
$f(\lambda)$ that vanishes as $\lambda\to\infty$
in the complex $\lambda$-plane has a matrix realization
in terms of three constant matrices $A,$ $B,$ $C$ as
$$f(\lambda)=-iC(\lambda I-iA)^{-1}B,\tag 3.2$$
where $I$ is the $p\times p$ identity
matrix, $A$ has size $p\times p,$
$B$ has size $p\times 1,$ and $C$
has size $1\times p$ for some
$p.$ We will refer
to $(A,B,C)$ as a matrix triplet of size $p.$ It is possible to
pad $A,$ $B,$ $C$ with zeros or it may be possible
to change them and increase or decrease the value of $p$
without changing $f(\lambda).$
The smallest
positive integer $p$ yielding $f(\lambda)$ gives
us a ``minimal" realization for $f(\lambda),$ and
it is known [8] that a minimal realization is unique
up to a similarity transformation.
Thus, without any loss of generality we can
always assume that our triplet
$(A,B,C)$ corresponds to a minimal
realization, and we will refer
to such a triplet as a minimal triplet. Note that
the poles of $f(\lambda)$
correspond to the eigenvalues of the matrix $(iA).$
By taking the Fourier transform of both sides of
(3.2), where the Fourier transform is defined as
$$\hat f(y):=\displaystyle\frac{1}{2\pi}\int_{-\infty}^\infty
d\lambda\,f(\lambda)\,e^{i\lambda y},$$
we obtain
$$\hat f(y)=Ce^{-Ay}B.\tag 3.3$$
We note that under the similarity transformation
$(A,B,C)\mapsto (S^{-1}AS,S^{-1}B,CS)$ for some
invertible matrix $S,$ the quantities
$f(\lambda)$ and $\hat f(y)$ remain unchanged.

Comparing (3.1) and (3.3) we see that they are closely related to each other.
As mentioned earlier, without loss of any generality we
assume that the real triplet $(A,B,C)$
in (3.1) corresponds to a minimal realization in (3.2).
For the time being, we will also assume
that all eigenvalues of $A$ in (3.1) have positive real parts. However,
in later sections we will relax the latter assumption and
choose our triplet in a less restrictive way, i.e. in the admissible
class $\Cal A$ defined in Section~4.

Let us use a dagger to denote the matrix adjoint (complex conjugation
and matrix transpose). Although the adjoint and the transpose
are equal to each other for real matrices, we will continue to use the dagger
notation even for the matrix transpose of real matrices so that
we can utilize the previous related results in [5,6]
obtained for the Zakharov-Shabat system and the
nonlinear Schr\"odinger equation.
Since $\Omega$ appearing in (3.1) is
a scalar we have $\Omega^\dagger=\Omega^*;$
thus, we get
$$\Omega(y,t)^*=B^\dagger e^{-A^\dagger y-(A^\dagger)^{-1}t/2} C^\dagger.$$

We note that when $\Omega$ is given by (3.1), the Marchenko equation
is exactly solvable by using linear algebra. This follows from
the separability property of the kernel, i.e.
$$\Omega(x+y,t)=Ce^{-Ax}e^{-Ay-A^{-1}t/2}B,\tag 3.4$$
indicating the separability in $x$ and $y;$ thus,
(3.4) allows us to try a solution to (2.4) in the form
$$K(x,y,t)=H(x,t)\,e^{-A^\dagger y-(A^\dagger)^{-1}t/2}C^\dagger.\tag 3.5$$
Using (3.5) in (2.4) we get
$$H(x,t)\,\Gamma(x,t)=B^\dagger e^{-A^\dagger x},\tag 3.6$$
or equivalently
$$H(x,t)=B^\dagger e^{-A^\dagger x}\Gamma(x,t)^{-1},\tag 3.7$$
where we have defined
$$\Gamma(x,t):=I+e^{-A^\dagger x-(A^\dagger)^{-1}t/2}Q
e^{-2Ax-A^{-1}t/2}Ne^{-A^\dagger x},\tag 3.8$$
with the constant $p\times p$ matrices $Q$ and $N$ defined as
$$Q:=\int_0^\infty ds\,e^{-A^\dagger s}C^\dagger C e^{-As},\qquad
N:=\int_0^\infty dr\,e^{-Ar}BB^\dagger e^{-A^\dagger r}.\tag 3.9$$
It is seen from (3.9) that $Q$ and $N$ are selfadjoint, i.e.
$$Q=Q^\dagger, \qquad N=N^\dagger.\tag 3.10$$
In fact, since the triplet $(A,B,C)$ is real,
the matrices $Q$ and $N$ are also real and hence they are
symmetric matrices.
Using (3.7) in (3.5) we obtain
$$K(x,y,t)=B^\dagger e^{-A^\dagger x}\Gamma(x,t)^{-1}\,
e^{-A^\dagger y-(A^\dagger)^{-1}t/2}C^\dagger,\tag 3.11$$
or equivalently
$$K(x,y,t)=B^\dagger F(x,t)^{-1}e^{-A^\dagger (y-x)}C^\dagger,\tag 3.12$$
where we have defined
$$F(x,t):=e^{\beta^\dagger}+Q
\,e^{-\beta}N,\tag 3.13$$
with the quantity $\beta$ defined as
$$\beta:=2Ax+\displaystyle\frac12\,A^{-1}t.\tag 3.14$$
 From (2.3) and (3.12) we see that
$$u(x,t)=-4\int_x^\infty dr\,B^\dagger F(r,t)^{-1}C^\dagger.\tag 3.15$$

The procedure described in (3.4)-(3.15) is exactly the same procedure used in
[5,6] with the only difference of using $A^{-1}t/2$ in the matrix
exponential in (3.15) instead
of $4iA^2t$ used in [5,6]. However, such a difference does not affect
the solution to the Marchenko integral equation at all
thanks to the fact that $A$ and $A^{-1}$ commute with each other.
In fact, the solution to the Marchenko equation is obtained the same way
if one replaces $A^{-1}/2$ by any function of the matrix
$A$ because such a matrix function commutes with $A.$

We will later prove that $F(x,t)$ given
in (3.13) is invertible on the entire $xt$-plane and
that $F(x,t)^{-1}\to 0$ exponentially as $x\to\pm\infty$ and hence
$u(x,t)$ given in (3.15)
is well defined on the entire $xt$-plane. We note that, as a result of
(2.6), the solution $K(x,y,t)$ to the Marchenko equation (2.4) is real
and hence (2.7) is satisfied. Hence, from
(2.3) we see that $u(x,t)$ is real valued, and by taking the adjoint of both sides of (3.15) we get
$$u(x,t)=-4\int_x^\infty dr\,C [F(r,t)^\dagger]^{-1}B.\tag 3.16$$

Instead of using (2.6) at the last stage, let us instead use
it from the very beginning when we solve the
Marchenko equation (2.4). Replacing $\Omega^*$ by $\Omega$ in the two occurrences
in (2.4), we can solve (2.4) in a similar way as in (3.4)-(3.15) and obtain
$$K(x,y,t)=C E(x,t)^{-1}e^{-A(y-x)} B,\tag 3.17$$
where we have defined
$$E(x,t):=e^{\beta}+P
\,e^{-\beta}P,\tag 3.18$$
with $\beta$ as in (3.14) and the constant matrix $P$ given by
$$P:=\int_0^\infty ds\,e^{-A s}B C e^{-As}.\tag 3.19$$
Thus, from (2.3) and (3.17) we obtain
$$u(x,t)=-4\int_x^\infty dr\,C E(r,t)^{-1}B.\tag 3.20$$
We will show in Section~5 that the two explicit solutions to the sine-Gordon equation
given by (3.15) and (3.20) are identical by proving that
$$E(x,t)=F(x,t)^\dagger.\tag 3.21$$

\vskip 10 pt
\noindent {\bf 4. EXACT SOLUTIONS USING THE LYAPUNOV EQUATIONS}
\vskip 3 pt

In Section~3 we have derived (3.15) and
(3.20) by assuming that we start with a real minimal triplet $(A,B,C)$ where
the eigenvalues of $A$ have positive real parts.
In this section we show that the explicit formula (3.15) for exact solutions
to the sine-Gordon equation
remains valid if the matrix triplet $(A,B,C)$ used to construct
such solutions is chosen in a larger class. Starting with a more arbitrary triplet
we will construct the matrix $F$ given in (3.13), where
the auxiliary matrices $Q$ and $N$ are no longer given by (3.9) but
obtained by uniquely solving the respective Lyapunov equations
$$A^\dagger Q+QA=C^\dagger C,\tag 4.1$$
$$AN+NA^\dagger=BB^\dagger .\tag 4.2$$

Many of the proofs in this section are similar to those obtained
earlier for the nonlinear Schr\"odinger equation [5,6] and hence
we will refer the reader to those references for the details of
some of the proofs.

\noindent {\bf Definition 4.1} {\it We say that the triplet $(A,B,C)$ of size $p$ belongs
to the admissible class $\Cal A$ if the following conditions are met:}
\item{(i)} {\it The matrices $A,$ $B,$ and $C$ are all real valued.}
\item{(ii)} {\it
The triplet $(A,B,C)$ corresponds to a minimal realization for
$f(\lambda)$ when that triplet is
used on the right hand side of (3.2).}
\item{(iii)} {\it None of the eigenvalues of
$A$ are purely imaginary and no two eigenvalues of
$A$ can occur symmetrically with respect to the imaginary axis in the complex
$\lambda$-plane.}

We note that, since $A$ is real valued, the condition stated in
(iii) is equivalent to
the condition that zero is not an eigenvalue of $A$ and that
no two eigenvalues of $A$ are located symmetrically with
respect to the origin in the complex plane.
Equivalently, (iii) can be stated as
$A$ and $(-A)$ not having any common eigenvalues.
We will say that
a triplet is admissible if it belongs to the
admissible class $\Cal A.$

Starting with a triplet $(A,B,C)$ in the admissible class
$\Cal A,$ we will obtain exact solutions to the
sine-Gordon equation as follows:

\item{(a)} Using $A,$ $B,$ $C$ as input, construct the auxiliary matrices
$Q$ and $N$ by solving the respective Lyapunov equations (4.1) and (4.2).
As the next theorem shows, the
solutions to (4.1) and (4.2) are unique and can be obtained as
$$Q=\displaystyle\frac{1}{2\pi}\int_{\gamma}
d\lambda\,(\lambda I+iA^\dagger)^{-1}C^\dagger C(\lambda I-iA)^{-1},\tag 4.3$$
$$N=\displaystyle\frac{1}{2\pi}\int_{\gamma}
d\lambda\,(\lambda I-iA)^{-1}BB^\dagger(\lambda I+iA^\dagger)^{-1},\tag 4.4$$
where $\gamma$ is any positively oriented simple closed contour enclosing
all eigenvalues of $(iA)$ and leaving out
all eigenvalues of $(-iA^\dagger).$ If all eigenvalues
of $A$ have positive real parts, then $Q$ and $N$ can also be evaluated as
in (3.9).

\item{(b)} Using the auxiliary matrices $Q$ and $N$ and the triplet $(A,B,C),$
form the matrix $F(x,t)$ as in (3.13)
and obtain the scalar $u(x,t)$ as in
(3.15), which becomes a solution to (1.1).

\noindent {\bf Theorem 4.2} {\it Consider any triplet $(A,B,C)$ belonging
to the admissible class $\Cal A$ described in Definition 4.1. Then:}

\item{(i)} {\it The Lyapunov equations (4.1) and (4.2) are uniquely solvable,
and their solutions are given by (4.3) and (4.4), respectively.}
\item{(ii)} {\it The constant matrices $Q$ and $N$
given in (4.3) and (4.4), respectively, are selfadjoint; i.e.
$Q^\dagger=Q$ and $N^\dagger=N.$ In fact, since the triplet
$(A,B,C)$ is real, the matrices
$Q$ and $N$ are also real. Furthermore, both $Q$ and $N$ are invertible.}

\item{(iii)} {\it The resulting matrix $F(x,t)$ formed as in (3.13) is real valued
and invertible on the entire $xt$-plane, and the function
$u(x,t)$ defined in (3.15) is a solution to the sine-Gordon equation
everywhere on the $xt$-plane. Moreover, $u(x,t)$ is
analytic on the entire $xt$-plane and $u_x(x,t)$ decays to zero
exponentially as $x\to\pm\infty$ at each fixed $t\in\bold R.$}

\noindent PROOF: The proof of (i) follows from Theorem~4.1 of
Section~4.1 of [20]. It is directly seen from (4.1) that
$Q^\dagger$ is also a solution whenever $Q$ is a solution, and
hence the uniqueness of the solution assures $Q=Q^\dagger.$
Similarly, as a result of the realness of the
triplet $(A,B,C),$ one can show that $Q^*$ is also
a solution to (4.1) and hence $Q=Q^*.$ The selfadjointness and realness of
$N$ are established the same way.
The invertibility of $Q$ and $N$ is a result of
the minimality of the triplet $(A,B,C)$
and a proof can be found in the proofs
of Theorems~3.2 and 3.3 of [5] by replacing (2.2)
of [5] with (3.13) in the current paper, completing the proof of (ii). From (3.13)
and (3.14)
it is seen that
the realness of the triplet $(A,B,C)$ and of $Q$ and $N$
implies the realness of $F.$
The proof of the invertibility
of $F$ is similar to the proof
of Proposition~4.1 (a)
of [5] and
the rest of the proof of (iii)
is obtained as in Theorem~3.2 (d) and (e)
of [5]. \qed

We will say that two triplets $(A,B,C)$ and $(\tilde A,\tilde B,\tilde C)$
are equivalent if they lead to the same $u(x,t)$ given in (3.15).
The next result shows that two admissible triplets are closely related to each other
and can always be transformed into each other.

\noindent {\bf Theorem 4.3} {\it
For any admissible triplet $(\tilde A,\tilde B,\tilde C),$
there corresponds an equivalent admissible
triplet $(A,B,C)$ in such a way that
all eigenvalues of $A$ have positive real parts.}

\noindent PROOF: The proof is similar to the proof of
Theorem~3.2 of [5], where
the triplet $(\tilde A,\tilde B,\tilde C)$ is expressed
explicitly when one starts with the triplet
$(A,B,C).$ Below we provide the explicit formulas
of constructing $(A,B,C)$ by starting with
$(\tilde A,\tilde B,\tilde C);$ i.e.,
by providing the inverse transformation
formulas for those given in [5].
Without loss of
any generality, we can assume that $(\tilde A,\tilde B,\tilde C)$ has the form
$$\tilde A=\bmatrix \tilde A_1&
0\\
\noalign{\medskip}
0&\tilde A_2\endbmatrix,\qquad \tilde B=\bmatrix \tilde B_1\\
\noalign{\medskip}
\tilde B_2\endbmatrix,\qquad \tilde C=\bmatrix \tilde C_1&\tilde C_2\endbmatrix,$$
where all eigenvalues of $\tilde A_1$ have positive real parts and
all eigenvalues of $\tilde A_2$ have negative real parts,
and for some
$0\le q\le p,$ the sizes of the matrices
$\tilde A_1,$  $\tilde A_2,$ $\tilde B_1,$  $\tilde B_2,$ $\tilde C_1,$  $\tilde C_2$ are
$q\times q,$ $(p-q)\times (p-q),$ $q\times 1,$
$(p-q)\times 1,$ $1\times q,$ and $1\times (p-q),$
respectively. We first construct the matrices
$\tilde Q$ and $\tilde N$ by solving the respective
Lyapunov equations
$$\cases \tilde Q\tilde A+\tilde A^\dagger \tilde Q=\tilde C^\dagger \tilde C,\\
\noalign{\medskip}
\tilde A\tilde N+\tilde N\tilde A^\dagger=\tilde B\tilde B^\dagger.\endcases$$
Writing $\tilde Q$ and $\tilde N$ in block matrix forms of
appropriate sizes as
$$\tilde Q=\bmatrix \tilde Q_1&\tilde Q_2\\
\noalign{\medskip}
\tilde Q_3&\tilde Q_4\endbmatrix,\qquad \tilde N=\bmatrix \tilde N_1&\tilde N_2\\
\noalign{\medskip}
\tilde N_3&\tilde N_4\endbmatrix,\tag 4.5$$
and, for appropriate block matrix sizes, by letting
$$A=\bmatrix A_1&
0\\
\noalign{\medskip}
0&A_2\endbmatrix,\qquad B=\bmatrix B_1\\
\noalign{\medskip}
B_2\endbmatrix,\qquad C=\bmatrix C_1&C_2\endbmatrix,\tag 4.6$$
we obtain
$$A_1=\tilde A_1,\qquad A_2=-\tilde A_2^\dagger,\qquad
B_1=\tilde B_1-\tilde N_2\tilde N_4^{-1}\tilde B_2,\qquad
B_2=\tilde N_4^{-1}\tilde B_2,\tag 4.7$$
$$C_1=\tilde C_1-\tilde C_2\tilde Q_4^{-1}\tilde Q_3,\qquad
C_2=\tilde C_2\tilde Q_4^{-1},\tag 4.8$$
yielding $(A,B,C)$ by starting with
$(\tilde A,\tilde B,\tilde C).$ \qed

When the triplet $(A,B,C)$ is decomposed as in (4.6), let us decompose
the corresponding solutions $Q$ and $N$ to the respective
Lyapunov equations (4.1) and (4.2), in an analogous manner to (4.5), as
$$Q=\bmatrix Q_1&Q_2\\
\noalign{\medskip}
Q_3&Q_4\endbmatrix,\qquad N=\bmatrix N_1&N_2\\
\noalign{\medskip}
N_3&N_4\endbmatrix.\tag 4.9$$
The relationship between (4.5) and (4.9) is summarized in the
following theorem.

\noindent {\bf Theorem 4.4} {\it
Under the transformation $(A,B,C)\mapsto (\tilde A,\tilde B,\tilde C)$
specified in Theorem~4.3, the quantities $Q,$ $N,$ $F,$ $E$ appearing
in (4.1), (4.2), (3.13), (3.18), respectively,
are transformed as}
$$(Q,N,F,E)\mapsto (\tilde Q,\tilde N,\tilde F,\tilde E),$$
{\it where}
$$\tilde Q=\bmatrix Q_1-Q_2 Q_4^{-1}Q_3&-Q_2Q_4^{-1}\\
\noalign{\medskip}
-Q_4^{-1}Q_3&-Q_4^{-1}\endbmatrix,\qquad N=\bmatrix N_1-N_2 N_4^{-1}N_3&-N_2N_4^{-1}\\
\noalign{\medskip}
-N_4^{-1}N_3&-N_4^{-1}\endbmatrix,$$
$$\tilde F=\bmatrix I&-Q_2Q_4^{-1}\\
\noalign{\medskip}
0&-Q_4^{-1}\endbmatrix F
\bmatrix I&0\\
\noalign{\medskip}
-N_4^{-1}N_3&-N_4^{-1}\endbmatrix,\tag 4.10$$
$$\tilde E=\bmatrix I&-N_2N_4^{-1}\\
\noalign{\medskip}
0&-N_4^{-1}\endbmatrix E
\bmatrix I&0\\
\noalign{\medskip}
-Q_4^{-1}Q_3&-Q_4^{-1}\endbmatrix.\tag 4.11$$

\noindent PROOF: The proof can be obtained in a similar manner to the
proof of Theorem~3.2 of [5] by using
$$\tilde A=\bmatrix A_1&0\\
\noalign{\medskip}
0&-A_2^\dagger\endbmatrix,\qquad
\tilde B=\bmatrix I&-N_2N_4^{-1}\\
\noalign{\medskip}
0&-N_4^{-1}\endbmatrix B,\qquad
\tilde C=C\bmatrix I&0\\
\noalign{\medskip}
-Q_4^{-1}Q_3&-Q_4^{-1}\endbmatrix,$$
corresponding to the transformation
specified in (4.7) and (4.8). \qed

As the following theorem shows, for an admissible triplet
$(A,B,C),$ there is no loss of generality in assuming that
all eigenvalues of $A$ have positive real parts and
$B$ has a special form consisting of zeros and ones.

\noindent {\bf Theorem 4.5} {\it
For any admissible triplet $(\tilde A,\tilde B,\tilde C),$
there correspond a special admissible
triplet $(A,B,C),$ where $A$ is in a Jordan canonical form
with each Jordan block containing a distinct eigenvalue having a positive real part, the entries of
$B$ consist of zeros
and ones, and $C$ has constant real entries. More specifically,
for some appropriate positive integer $m$ we have
}
$$A=\bmatrix A_1&
0&\cdots&0\\
0& A_2&\cdots&0\\
\vdots&\vdots &\ddots&\vdots\\
0&0&\cdots&A_m\endbmatrix,
\qquad
B=\bmatrix
B_1\\
B_2\\
\vdots\\
B_m\endbmatrix,\qquad
C=\bmatrix C_1&C_2& \cdots
&
C_m\endbmatrix,\tag 4.12$$
{\it where in the case of a real (positive) eigenvalue $\omega_j$
of $A_j$ the corresponding blocks are given by}
$$C_j:=\bmatrix c_{jn_j}& \cdots
&
c_{j 2}&
c_{j 1}\endbmatrix,\tag 4.13$$
$$A_j:=\bmatrix \omega_j&
-1&0&\cdots&0&0\\
0& \omega_j& -1&\cdots&0&0\\
0&0&\omega_j& \cdots&0&0\\
\vdots&\vdots &\vdots &\ddots&\vdots&\vdots\\
0&0&0&\cdots&\omega_j&-1\\
0&0&0&\cdots&0&\omega_j\endbmatrix,
\qquad
B_j:=\bmatrix 0\\
\vdots\\
0\\
1\endbmatrix,\tag 4.14$$
{\it with $A_j$ having size $n_j\times n_j,$ $B_j$ size $n_j\times 1,$
$C_j$ size $1\times n_j,$  and
the constant $c_{jn_j}$ is nonzero.
In the case of complex eigenvalues, which must appear
in pairs as $\alpha_j\pm i\beta_j$ with $\alpha_j>0,$ the corresponding
blocks are given by}
$$C_j:=\bmatrix \gamma_{j n_j}& \epsilon_{j n_j}&\dots
&
\gamma_{j 1}& \epsilon_{j 1}\endbmatrix,\tag 4.15$$
$$A_j:=\bmatrix \Lambda_j&
-I_2&0&\dots&0&0\\
0& \Lambda_j& -I_2&\dots&0&0\\
0&0&\Lambda_j&\dots&0&0\\
\vdots&\vdots &\vdots &\ddots&\vdots&\vdots\\
0&0&0&\dots&\Lambda_j&-I_2\\
0&0&0&\dots&0&\Lambda_j\endbmatrix
,\quad
B_j:=\bmatrix 0\\
\vdots\\
0\\
1\endbmatrix,\tag 4.16$$
{\it where $\gamma_{js}$ and $\epsilon_{js}$ for
$s=1,\dots,n_j$ are
real constants with $(\gamma_{jn_j}^2+\epsilon_{jn_j}^2)>0,$ $I_2$ denotes the
$2\times 2$ unit matrix, each column vector
$B_j$ has $2n_j$ components, each $A_j$
has size $2n_j\times 2n_j,$
and each $2\times 2$
matrix $\Lambda_j$ is defined as}
$$\Lambda_j:=\bmatrix \alpha_j&\beta_j
\\
\noalign{\medskip}
-\beta_j& \alpha_j\endbmatrix.\tag 4.17$$

\noindent PROOF: The real triplet $(A,B,C)$ can be chosen as described
in Section~3 of [7]. \qed

\vskip 10 pt
\noindent {\bf 5. EXACT SOLUTIONS USING THE SYLVESTER EQUATION}
\vskip 3 pt

In Section 3, starting from a minimal triplet $(A,B,C)$
with all eigenvalues of $A$ having positive real parts, we have obtained the exact
solution formula (3.20) to the sine-Gordon equation
by constructing the matrix $E(x,t)$ in (3.18) with the help
of the auxiliary matrix $P$ in (3.19). In
this section we show that the explicit formula (3.20) for exact solutions
to the sine-Gordon equation remains valid
if the matrix triplet $(A,B,C)$ used to construct
such solutions comes from a larger class, namely from the
admissible class $\Cal A$ specified in Definition 4.1.

Starting with any triplet $(A,B,C)$
in the admissible class $\Cal A,$ we obtain exact solutions
to the sine-Gordon equation as follows:

\item{(a)} Using $A,$ $B,$ $C$ as input, construct the auxiliary matrix
$P$ by solving the Sylvester equation
$$AP+PA=BC.\tag 5.1$$
The unique solution to (5.1) can be obtained as
$$P=\displaystyle\frac{1}{2\pi}\int_{\gamma}
d\lambda\,(\lambda I-iA)^{-1}BC(\lambda I+iA)^{-1},\tag 5.2$$
where $\gamma$ is any positively oriented simple closed contour enclosing
all eigenvalues of $(iA)$ and
leaving out
all eigenvalues of $(-iA).$
If all eigenvalues of $A$ have positive real parts, then
$P$ can be evaluated as in (3.19).

\item{(b)} Using the auxiliary matrix $P$ and the triplet $(A,B,C),$
form the matrix $E(x,t)$ as in (3.18)
and then form the scalar $u(x,t)$ as in
(3.20).

\noindent {\bf Theorem 5.1} {\it Consider any triplet $(A,B,C)$ belonging
to the admissible class $\Cal A$ described in Definition 4.1. Then,
the Sylvester equation (5.1) is uniquely solvable, and
its solution is given by (5.2). Furthermore, that solution is real valued.}

\noindent PROOF: The unique solvability of (5.1) is already known [20]. For the benefit of the reader
we outline the steps below. From (5.1) we get
$$-(\lambda I-iA)P+P(\lambda I+iA)=iBC,$$
or equivalently
$$-P(\lambda I+iA)^{-1}+(\lambda I-iA)^{-1}P
=i(\lambda I-iA)^{-1}BC(\lambda I+iA)^{-1}.
\tag 5.3$$
Dividing both sides of (5.3) by $(2\pi)$ and then integrating along $\gamma,$
and using
$$\displaystyle\frac{1}{2\pi i}\int_\gamma d\lambda\,(\lambda I-iA)^{-1}=I,
\qquad
\displaystyle\frac{1}{2\pi i}\int_\gamma d\lambda\,(\lambda I+iA)^{-1}=0,$$
we obtain (5.2) as the unique solution to (5.1).
Since the admissible triplet $(A,B,C)$ is real,
by taking complex conjugate of both sides of
(5.1) we see that $P^*$ also solves (5.1). From the uniqueness
of the solution to (5.1), it then follows the $P^*=P.$ \qed

Next we show that, for any triplet
$(A,B,C)$ in our admissible class $\Cal A,$
there is a close relationship between the matrix $P$
given in (5.2) and
the matrices
$Q$ and $N$ appearing in (4.3) and (4.4), respectively.

\noindent {\bf Theorem 5.2} {\it Let
the triplet $(A,B,C)$ of size $p$ belong to the admissible class
specified in Definition 4.1. Then
the solution $P$ to the Sylvester equation (5.1) and
the solutions $Q$ and $N$ to the respective
Lyapunov equations (4.1) and (4.2) satisfy}
$$NQ=P^2.\tag 5.4$$

\noindent PROOF: Note that (5.4) is valid when the matrix $A$ in the triplet
is diagonal. To see this, note that the use of the triplet $(A,B,C)$ with
$$A=\text{diag}\{a_1,\cdots,a_p\},\qquad B=\bmatrix b_1\\
\vdots\\
b_p\endbmatrix,\qquad
C=\bmatrix c_1&\cdots& c_p\endbmatrix,$$
in (4.1), (4.2), and (5.1) yields
$$P_{jk}=\displaystyle\frac{b_jc_k}{a_j+a_k},\qquad
Q_{jk}=\displaystyle\frac{c_jc_k}{a_j+a_k},\qquad
N_{jk}=\displaystyle\frac{b_jb_k}{a_j+a_k},$$
where the subscript $jk$ denotes the $(j,k)$ entry
of the relevant matrix. Hence,
$$(NQ)_{jk}=\displaystyle\sum_{s=1}^p \displaystyle\frac{b_jb_sc_sc_k}{(a_j+a_s)(a_s+a_k)},
\quad
(P^2)_{jk}=\displaystyle\sum_{s=1}^p \displaystyle\frac{b_jc_sb_sc_k}{(a_j+a_s)(a_s+a_k)},$$
establishing (5.4). Next, let us assume that $A$ is not diagonal but
diagonalizable through a real-valued invertible matrix
$S$ so that $\tilde A=S^{-1}AS$ and $\tilde A$
is diagonal. Then, under the transformation
$$(A,B,C)\mapsto(\tilde A,\tilde B,\tilde C)=(S^{-1}AS,S^{-1}B,CS),$$
we get
$$(Q,N,P)\mapsto (\tilde Q,\tilde N,\tilde P)=(S^\dagger QS,S^{-1}N(S^\dagger)^{-1},S^{-1}PS),$$
where $\tilde Q,$ $\tilde N,$ and $\tilde P$ satisfy (4.1), (4.2), and (5.1),
respectively, when $(A,B,C)$ is replaced with $(\tilde A,\tilde B,\tilde C)$
in those three equations. We note that $(\tilde A,\tilde B,\tilde C)$ is an
admissible triplet when $(A,B,C)$ is admissible because the eigenvalues
of $A$ and $\tilde A$ coincide.
Since $\tilde A$ is diagonal, we already have
$\tilde N\tilde Q=\tilde P^2,$ which easily reduces to
$NQ=P^2$ given in (5.4).
In case $A$ is not diagonalizable, we proceed as follows. There exists
a sequence of admissible triplets $(A_k,B,C)$ converging to
$(A,B,C)$ as $k\to+\infty$ such that each $A_k$ is diagonalizable. Let
the triplet $(Q_k,N_k,P_k)$ correspond to the solutions
to (4.1), (4.2), and (5.1),
respectively, when $(A,B,C)$ is replaced with $(A_k,B,C)$
in those three equations.
We then have
$N_kQ_k=P_k^2,$ and hence $(Q_k,N_k,P_k)\to (Q,N,P)$
yields (5.4).
Note that we have used the stability of solutions
to (4.1), (4.2), and (5.1). In fact, that stability directly follows from
the unique solvability of the matrix equations
(4.1), (4.2), (5.1) and the fact that their unique
solvability is preserved under a small perturbation of $A.$ \qed

\noindent {\bf Theorem 5.3} {\it Let
the triplet $(A,B,C)$ belong to the admissible class
specified in Definition 4.1. Then,
the solution $P$ to the Sylvester equation (5.1) and
the solutions $Q$ and $N$ to the respective
Lyapunov equations (4.1) and (4.2) satisfy}
$$N(A^\dagger)^jQ=PA^j P,\qquad j=0,\pm 1,\pm 2,\dots.\tag 5.5$$

\noindent PROOF: Under the transformation
$$(A,B,C)\mapsto(\tilde A,\tilde B,\tilde C)=(A,A^j B,C),$$
we get
$$(Q,N,P)\mapsto (\tilde Q,\tilde N,\tilde P)=(Q,A^j N(A^\dagger)^j,A^j P),$$
where $\tilde Q,$ $\tilde N,$ and $\tilde P$ satisfy (4.1), (4.2), and (5.1),
respectively, when $(A,B,C)$ is replaced with $(\tilde A,\tilde B,\tilde C)$
in those three equations. Since $(\tilde A,\tilde B,\tilde C)$ is also
admissible, (5.4) implies that
$\tilde N\tilde Q=\tilde P^2,$ which yields
(5.5) after a minor simplification. \qed

Next, given any admissible triplet $(A,B,C),$ we prove that
the corresponding solution $P$ to (5.1) is invertible and that
the matrix $E(x,t)$ given in (3.18) is invertible and that
(3.21) holds everywhere on the $xt$-plane.

\noindent {\bf Theorem 5.4} {\it Let
the triplet $(A,B,C)$ belong to the admissible class
specified in Definition 4.1, and let the matrices
$Q,$ $N,$ $P$ be the corresponding
solutions to (4.1), (4.2), and (5.2), respectively.
Then:}

\item{(i)} {\it The matrix $P$
is invertible.}

\item{(ii)} {\it The matrices $F$ and $E$ given in (3.13) and (3.18), respectively,
are real valued and satisfy
(3.21).}

\item{(iii)} {\it The matrix $E(x,t)$ is
invertible on the entire $xt$-plane.}

\noindent PROOF: The invertibility of $P$ follows from (5.4) and the fact that
both $Q$ and $N$ are invertible, as stated in Theorem~4.2 (ii);
thus, (i) is established. To prove (ii) we proceed as follows.
The real-valuedness of
$F$ has already been established in Theorem~4.2 (iii). From (3.18)
it is seen that the real-valuedness of
the triplet $(A,B,C)$ and of $P$ implies
that $E$ is real valued. From (3.13), (3.14), and (3.18) we see that
(3.21) holds if and only if we have
$$N e^{-\beta^\dagger}Q=P e^{-\beta}P,
\tag 5.6$$
where we have already used $N^\dagger=N$ and $Q^\dagger=Q,$ as
established in Theorem~4.2 (ii). Since (5.5) implies
$$N(-\beta^\dagger)^jQ=P(-\beta)^jP,\qquad j=0,1,2,\dots,$$
we see that (5.6) holds.
Having established (3.21), the invertibility of
$E(x,t)$ on the entire $xt$-plane
follows from the invertibility of
$F(x,t),$ which has been established in Theorem~4.2 (iii). \qed

Next, we show that the explicit formulas (3.15),
(3.16), and (3.20) are all equivalent to each other.

\noindent {\bf Theorem 5.5} {\it Consider any triplet $(A,B,C)$ belonging
to the admissible class $\Cal A$ described in Definition 4.1. Then:}

\item{(i)} {\it The explicit formulas (3.15),
(3.16), and (3.20) yield equivalent exact solutions to the sine-Gordon equation (1.1)
everywhere on the entire $xt$-plane.}

\item{(ii)} {\it The equivalent solution $u(x,t)$ given in
(3.15), (3.16), and (3.20) is analytic on the entire $xt$-plane,
and $u_x(x,t)$ decays to zero
exponentially as $x\to\pm\infty$ at each fixed $t\in\bold
R.$}

\noindent PROOF: Because $u(x,t)$
is real and scalar valued, we already have the equivalence
of (3.15) and (3.16). The equivalence of (3.16) and (3.20)
follows from (3.21). We then have (ii) as a consequence of
Theorem~4.2 (iii). \qed

\vskip 10 pt
\noindent {\bf 6. FURTHER EQUIVALENT FORMS FOR EXACT SOLUTIONS}
\vskip 3 pt

In Theorem~5.5 we have shown that the exact solutions given by
the explicit
formulas (3.15), (3.16), and (3.20) are equivalent.
In this section we show that our exact solutions can be written in various
other equivalent forms. We first present two propositions that will be useful in later sections.

\noindent {\bf Proposition 6.1} {\it If $(A,B,C)$ is admissible, then
the quantities $F^{-1}$ and $E^{-1},$ appearing in
(3.13) and (3.18), respectively, vanish
exponentially as $x\to\pm\infty.$}

\noindent PROOF: It is sufficient to give the proof when the eigenvalues of
$A$ have all positive real parts because, as seen from (4.10) and (4.11), the same result also holds when some or all eigenvalues of $A$ have negative
real parts. When the eigenvalues of $A$ have positive real parts, from (3.13) we get
$$F^{-1}=e^{-\beta^\dagger/2}[I+e^{-\beta^\dagger/2}Qe^{-\beta}
Ne^{-\beta^\dagger/2}]^{-1}e^{-\beta^\dagger/2},\tag 6.1$$
where the invertibility
of $Q$ and $N$ is guaranteed by Theorem~4.2 (ii).
Hence, (6.1) implies that $F^{-1}\to 0$
exponentially as $x\to+\infty.$ From
(3.21) and the realness of $E$ and $F$ we also get
$E^{-1}\to 0$ exponentially as $x\to+\infty.$
To obtain the asymptotics as $x\to-\infty,$ we proceed as follows.
 From (3.13) we obtain
$$Q^{-1}F N^{-1}= e^{-\beta/2}
[I+e^{\beta/2}Q^{-1}e^{\beta^\dagger}N^{-1}e^{\beta/2}]e^{-\beta/2},$$
and hence
$$F^{-1}=N^{-1}e^{\beta/2}
[I+e^{\beta/2}Q^{-1}e^{\beta^\dagger}N^{-1}e^{\beta/2}]^{-1}
e^{\beta/2}Q^{-1},$$
and thus $F^{-1}\to 0$ exponentially as $x\to-\infty.$ From
(3.21) and the realness of $E$ and $F$ we also get
$E^{-1}\to 0$ exponentially as $x\to-\infty.$ \qed

\noindent {\bf Proposition 6.2} {\it The quantity $E(x,t)$ defined
in (3.18) satisfies}
$$E_x=2AE-2BCe^{-\beta}P,
\qquad Ee^{-\beta}P=Pe^{-\beta}E,\qquad
e^{\beta}P^{-1}E=P+e^{\beta}P^{-1}e^{\beta}.\tag 6.2$$
{\it If $(A,B,C)$ is admissible and all eigenvalues of
$A$ have positive real parts, then
$E^{-1}Pe^{-\beta}\to P^{-1}$ exponentially
as $x\to-\infty.$}

\noindent PROOF: We obtain the first
equality (6.2) by taking the $x$-derivative of (3.18) and by
using (5.1). The second equality can be verified directly
by using (3.18) in it.
The third equality is obtained by a direct premultiplication from (3.18). The limit
as $x\to-\infty$ is seen from the last equality in
(6.2) with the help of (3.14). \qed

Let us start with a triplet $(A,B,C)$ of size $p$ belonging to
the admissible class specified in Definition 4.1. Letting
$$M(x,t):=e^{-\beta/2}Pe^{-\beta/2},\tag 6.3$$
where $\beta$ as in (3.14) and
$P$ is the unique solution to the Sylvester equation (5.1),
we can write (3.18) also as
$$E(x,t)=e^{\beta/2}\Lambda
e^{\beta/2},\tag 6.4$$
where we have defined
$$\Lambda(x,t):=I+[M(x,t)]^2.\tag 6.5$$
Using (5.1) in (6.3), we see that the $x$-derivative of
$M(x,t)$ is given by
$$M_x(x,t)=-e^{-\beta/2}BCe^{-\beta/2}.\tag 6.6$$

\noindent {\bf Proposition 6.3} {\it The eigenvalues of the
matrix $M$ defined
in (6.3) cannot occur on the imaginary axis in the complex plane.
Furthermore, the matrices $(I-iM)$ and $(I+iM)$ are invertible
on the entire $xt$-plane.}

\noindent PROOF: From (6.4) and (6.5) we see that
$$(I-iM)(I+iM)=e^{-\beta/2}E
e^{-\beta/2},$$
and by Theorem~5.4 (iii) the matrix $E$ is invertible
on the entire $xt$-plane. Thus, both $(I-iM)$ and $(I+iM)$ are invertible,
and consequently $M$ cannot have eigenvalues $\pm i.$
For any real, nonzero $c,$ consider the transformation
$(A,B,C)\mapsto (A,cB,cC)$ of an admissible triple $(A,B,C).$
The resulting triple is also admissible, and as seen from
(5.1) and (6.3) we have
$(P,M,I+M^2)\mapsto (c^2P,c^2M,I+c^4M^2).$ Thus,
$M$ cannot have any purely imaginary eigenvalues.
Since $P$ is known to be invertible by Theorem~5.4 (i), as
seen from (6.3) the matrix $M$ is invertible
on the entire $xt$-plane and hence cannot have
zero as its eigenvalue. \qed

\noindent {\bf Theorem 6.4} {\it The solution to the sine-Gordon equation
given in the equivalent forms (3.15), (3.16), and (3.20) can also be written as}
$$u(x,t)=-4 \text{Tr}[\tan^{-1} M(x,t)],\tag 6.7$$
$$u(x,t)=2i\log\left(
\displaystyle\frac{\det(I+iM(x,t))}
{\det(I-iM(x,t))}
\right),\tag 6.8$$
$$u(x,t)=4\tan^{-1}\left(i\,\displaystyle\frac{\det(I+iM(x,t))-\det(I-iM(x,t))}{\det(I+iM(x,t))
+\det(I-iM(x,t))}\right),\tag 6.9$$
{\it where $M$ is the matrix defined in (6.3)
and $\text{Tr}$ denotes the matrix trace (the sum of diagonal entries).}

\noindent PROOF: Let us note that the equivalence of (6.8) and (6.9) follows from the
second equality in (1.5) by using
$z=\det(I+iM)/\det(I-iM)$ there. To show the equivalence of (6.7) and (6.8),
we use the matrix identity
$$\tan^{-1}M=\displaystyle\frac{1}{2i}
\log\left((I+iM)(I-iM)^{-1}\right),$$
which is closely related to the first identity in (1.5), and the matrix identity
$$\text{Tr}[\log z]=\log\det z,$$
with the invertible matrix $z=(I+iM)(I-iM)^{-1}.$ Thus, we have established the
equivalence of (6.7), (6.8), and (6.9). We will complete the proof by showing that
(3.20) is equivalent to (6.7).
Using the fact that for any $m\times n$ matrix
$\alpha$ and any $n\times m$ matrix $\gamma$ we have $$\text{Tr}[\alpha\gamma]=
\text{Tr}[\gamma\alpha],\tag 6.10$$
 from (6.4)-(6.6) we get
$$-4CE^{-1}B=4\text{Tr}[M_x(I+M^2)^{-1}].\tag 6.11$$
By Proposition~6.1 we know that $E^{-1}$ vanishes exponentially as $x\to+\infty.$
Hence, with the help of (6.11) we see that
we can write (3.20) as
$$u(x,t)=4\text{Tr}\left[
\int_x^\infty dr\,M_r(r,t)[I+M(r,t)\,M(r,t)]^{-1}\right],$$
which yields (6.7). \qed

\noindent {\bf Theorem 6.5} {\it The solution to the sine-Gordon equation
given in the equivalent forms (3.15), (3.16), (3.20),
(6.7)-(6.9) can also be written as}
$$u(x,t)=-4\sum_{j=1}^p\tan^{-1}\kappa_j(x,t),\tag 6.12$$
{\it where the scalar functions
$\kappa_j(x,t)$ correspond to the eigenvalues
of the matrix
$M(x,t)$ defined in (6.3) and the repeated eigenvalues
are allowed in the summation.}

\noindent PROOF: At a fixed $(x,t)$-value, using the matrix identity
$$\text{Tr}[M(x,t)^s]=\sum_{j=1}^p [\kappa_j(x,t)]^s,\qquad s=1,2,3,\dots,$$
for large $|z|$ values in the complex $z$-plane we obtain
$$\text{Tr}[(z I-M)^{-1}]=\sum_{s=0}^\infty
z^{-s-1}\text{Tr}[M^s]=
\sum_{s=0}^\infty \sum_{j=1}^p z^{-s-1} \kappa_j^s
=\sum_{j=1}^p (z-\kappa_j)^{-1},\tag 6.13$$
where we dropped the arguments of $M$ and $\kappa_j$ for simplicity.
Choosing the contour $\gamma$ as in (1.6) so that each
eigenvalue $\kappa_j(x,t)$ is encircled exactly once in the positive
direction, we can extend (6.13) to $z\in\gamma$
by an analytic continuation with respect to $z.$ Using (6.12) in
(1.6), we
then obtain
$$\displaystyle\frac{1}{2\pi i}\int_\gamma
dz\,[\tan^{-1}z]\,
\text{Tr}[(z I-M)^{-1}]=
\sum_{j=1}^p
\displaystyle\frac{1}{2\pi i}\int_\gamma
dz\,[\tan^{-1}z]\,
(z-\kappa_j)^{-1},$$
or equivalently
$$\text{Tr}[\tan^{-1} M(x,t)]=\sum_{j=1}^p
\tan^{-1}\kappa_j(x,t),$$
which yields (6.12) in view of (6.7). \qed

Let us note that the equivalence of (6.7)-(6.9), and (6.12) implies that
one can
replace $M$ by its Jordan canonical form in any of those four expressions
without changing the value of $u(x,t).$ This follows from the
fact that $u(x,t)$ in (6.8)
remains unchanged if $M$ is replaced by its Jordan canonical form
and is confirmed in (6.12) by the fact that the eigenvalues
remain unchanged under a similarity transformation on a matrix.

The next result shows that we can write our explicit solution given in (6.12) yet another
equivalent form, which is expressed in terms of the coefficients in the
characteristic polynomial of the
matrix $M(x,t)$ given in (6.8). Let that characteristic polynomial be given by
$$\det\left(z I-M(x,t)\right)=\prod_{j=1}^p \left[z-\kappa_j(x,t)\right]
=\sum_{j=0}^p
(-1)^j\sigma_j(x,t)\, z^{p-j},$$
where the coefficients $\sigma_j(x,t)$ can be written in terms of the eigenvalues
$\kappa_j(x,t)$ as
$$\sigma_0=1,\qquad
\sigma_1=\sum_{j=1}^p \kappa_j,\qquad
\sigma_2=\sum_{1\le j<k\le p}^p \kappa_j\kappa_k,\qquad
\dots\quad,\quad \sigma_p=\kappa_1\cdots\kappa_p,\tag 6.14
$$
where we have dropped the arguments and have written $\kappa_j$ and
$\sigma_j$ for $\kappa_j(x,t)$ and $\sigma_j(x,t),$ respectively, for
simplicity.

\noindent {\bf Theorem 6.6} {\it The solution to the sine-Gordon equation
given in the equivalent forms (3.15), (3.16), (3.20),
(6.7)-(6.9), and (6.12) can also be written as}
$$u(x,t)=-4\tan^{-1}\left(
\displaystyle\frac{\displaystyle\sum_{s=0}^{\lfloor (p-1)/2\rfloor} (-1)^s\sigma_{2s+1}(x,t)}
{\displaystyle\sum_{s=0}^{\lfloor p/2\rfloor} (-1)^s\sigma_{2s+1}(x,t)}
\right)
,\tag 6.15$$
{\it where $\lfloor j\rfloor$ denotes the greatest integer
function of $j$ and
the quantities $\sigma_j(x,t)$ are those given in
(6.14).}

\noindent PROOF: When $p=2,$ by letting
$\eta_j:=\tan^{-1}\kappa_j(x,t)$ and using the addition formula for the tangent function,
we obtain
$$\tan(\eta_1+\eta_2)=\frac{\tan\eta_1+\tan\eta_2}{1-(\tan\eta_1)(\tan\eta_2)}
=\frac{\kappa_1+\kappa_2}{1-\kappa_1\kappa_2}=\frac{\sigma_1}{\sigma_0-\sigma_2},\tag 6.16$$
and hence the application of the inverse tangent function on both sides of (6.16)
yields (6.15). For larger values of $p,$ we proceed by induction with
respect to $p$ and by the further use of
the addition formula for the tangent function. \qed

\vskip 10 pt
\noindent {\bf 7. FURTHER PROPERTIES OF OUR EXACT SOLUTIONS}
\vskip 3 pt

In this section we derive
an explicit expression, in terms of a matrix triplet, for the square of the
spatial derivative of our exact solutions to (1.1) and analyze
further properties of such solutions.

\noindent {\bf Theorem 7.1} {\it If $(A,B,C)$ is admissible, then
the solution to the sine-Gordon equation
given in the equivalent forms (3.15), (3.16), (3.20),
(6.7)-(6.9), (6.12), and (6.15) satisfy}
$$[u_x(x,t)]^2=
\text{Tr}[(\Lambda^{-1}\Lambda_x)_x]
=\text{Tr}[(E^{-1}E_x)_x]=
\text{Tr}[(F^{-1}F_x)_x]
,\tag 7.1$$
{\it where $\Lambda,$  $E,$ and $F$ are the quantities appearing
in (6.5), (3.18), and (3.13), respectively. Consequently,
we have}
$$[u_x(x,t)]^2
=
\displaystyle\frac{\partial^2 \log(\det \Lambda(x,t))}{\partial x^2}
=
\displaystyle\frac{\partial^2 \log(\det E(x,t))}{\partial x^2}=
\displaystyle\frac{\partial^2 \log(\det F(x,t))}{\partial x^2}
.\tag 7.2$$

\noindent PROOF: Let us use the notation of
Theorems~4.3 and 4.4 and use a tilde to denote the quantities
associated with the triplet $(\tilde A,\tilde B,\tilde C),$
where some or all eigenvalues of $\tilde A$ have negative
real parts. Because of the equivalence stated in Theorems~4.3 and 4.4, we can convert the starting triplet
$(\tilde A,\tilde B,\tilde C)$ into an admissible triplet
$(A,B,C)$ where the matrix
$A$ has eigenvalues with positive
real parts. We will first establish (7.1) and (7.2) for
the quantities associated with the triplet
$(A,B,C)$ and then show that those formulas
remain valid when we use $(\tilde A,\tilde B,\tilde C)$ as the input triplet.
We exploit the connection
between (1.4) and the Zakharov-Shabat system
given in (2.1) of [6], where
$q=-iu_x/2$ and $u$ is real valued.
From (2.4) and (2.10) of [6] we see that
$$[u_x(x,t)]^2=8\displaystyle\frac{\partial G(x,x,t)}{\partial x},
\qquad \int_x^\infty  dr\,[u_r(r,t)]^2=-8
G(x,x,t),\tag 7.3$$
where we have
$$G(x,y,t)=-\int_x^\infty dr K(x,r,t)^*\,\Omega(r+y,t)^*,\tag 7.4$$
with $K(x,y,t)$ given in the equivalent forms (3.12) or (3.17),
and $\Omega(r+y,t)$ given in (3.4). Since our triplet $(A,B,C)$ is real, both
$K$ and $\Omega$ are real valued and we can ignore
the complex conjugations in the integrand in (7.4). Thus, we get
$$G(x,y,t)=-CE(x,t)^{-1}\int_x^\infty dr\,
e^{-A(r-x)}BC e^{-A(r+y)-A^{-1}t/2}B,\tag 7.5$$
which is evaluated with the help of (3.19) as
$$G(x,y,t)=-CE(x,t)^{-1}Pe^{-\beta}e^{-A(y-x)}B,\tag 7.6$$
where $\beta$ is the quantity in (3.14). Omitting the arguments
$(x,t)$ and using (7.6) in (7.3) we get
$$u_x^2=-8[CE^{-1}Pe^{-\beta}B]_x.\tag 7.7$$
Using (6.3) and (6.4) in (7.7) we obtain
$$u_x^2=-8[Ce^{-\beta/2}\Lambda^{-1}Me^{-\beta/2}B]_x,\tag 7.8$$
where $M$ is the quantity defined in (6.3). With the help of (6.10)
we write (7.8) as
$$u_x^2=-8\text{Tr}[e^{-\beta/2}BCe^{-\beta/2}\Lambda^{-1}M]_x,$$
or equivalently, after using (6.6), we get
$$u_x^2=-8\text{Tr}[M_x\Lambda^{-1}M]_x,$$
Using (6.5) and the fact that $M$ and $\Lambda^{-1}$ commute,
we obtain the first equality in (7.1).
With the help of (6.4) we obtain
$$\Lambda_x=-A\Lambda-\Lambda A+e^{-\beta/2}E_xe^{-\beta/2},$$
$$\Lambda^{-1}\Lambda_x=
-\Lambda^{-1}A\Lambda-A+\Lambda^{-1}e^{-\beta/2}E_xe^{-\beta/2},
\tag 7.9$$
and hence using (6.4) and (6.10), from (7.9) we obtain
$$\text{Tr}[\Lambda^{-1}\Lambda_x]=-2\text{Tr}[A]+
\text{Tr}[E^{-1}E_x],\tag 7.10$$
establishing the second equality in (7.1). With the help
of (3.21) and the fact that $E$ and $F$ are real valued,
we establish the third equality in (7.1). Using the matrix
identity
$$\text{Tr}[\alpha^{-1}\alpha_x]=\displaystyle\frac{1}
{\det\alpha}\displaystyle\frac{\partial\det\alpha}
{\partial x}=\displaystyle\frac{\partial \log(\det\alpha)}{\partial x},$$
we write (7.1) in the equivalent form of (7.2).
Now, if we use $(\tilde A,\tilde B,\tilde C)$ instead of
$(A,B,C),$ we see from (4.11) that,
for some constant invertible matrices
$Y$ and $Z,$ we have
$$\tilde E=YEZ,\qquad
\tilde E^{-1}=Z^{-1}E^{-1}Y^{-1},
\qquad
\tilde E_x=Y E_x Z,\tag 7.11$$
and hence, with the help of (6.10) and (7.11) we get
$$\text{Tr}[\tilde E^{-1}\tilde E_x]=\text{Tr}[E^{-1}E_x].\tag 7.12$$
Similarly, (4.10) yields
$$\tilde F=Z^\dagger F Y^\dagger,\qquad
\tilde F^{-1}=(Y^\dagger)^{-1}F^{-1}(Z^\dagger)^{-1},
\qquad
\tilde F_x=Z^\dagger F_x Y^\dagger,$$
which yields
$$\text{Tr}[\tilde F^{-1}\tilde F_x]=\text{Tr}[F^{-1}F_x].\tag 7.13$$
Note that from (7.10) and (7.12) we get
$$\text{Tr}[\tilde \Lambda^{-1}\tilde \Lambda_x]+2\,\text{Tr}[\tilde A]
=\text{Tr}[\Lambda^{-1}\Lambda_x]+2\,\text{Tr}[A].\tag 7.14$$
Thus, by taking the $x$-derivatives of both sides in
(7.12), (7.13), and (7.14), we establish
(7.1) and (7.2) without any restriction on the
sign of the real parts of the eigenvalues of $A.$ \qed

Next, we show that the proof of Theorem~7.1 can be obtained directly without
using (7.3)-(7.6). For this purpose, it is sufficient for us to show that
(7.7) can directly be derived from (3.20).

\noindent {\bf Proposition 7.2} {\it The equality in (3.20) implies (7.7), i.e. we have}
$$-8(CE^{-1}PE^{-\beta}B)_x=16CE^{-1}BCE^{-1}B.\tag 7.15$$

\noindent PROOF: We directly evaluate the left hand side
of (7.15) by taking the $x$-derivative of
$E^{-1}Pe^{-\beta}.$ We simplify the resulting expression
by using the first two equalities given in (6.2), and we obtain
the right hand side in (7.15). \qed

The next result shows
that $u(-\infty,t)$ must be an integer
multiple of $2\pi.$
In fact, we have $u(-\infty,t)=2\pi j,$ where
$j\in\{-p,-p+1,\dots,0,\dots,
p-1,p\},$ with $p$ denoting the size of the triplet $(A,B,C)$ used to
construct our exact solutions.

\noindent {\bf Theorem 7.3} {\it If $(A,B,C)$ is admissible
and the eigenvalues of $A$ have positive real parts, then
the solution to the sine-Gordon equation
given in the equivalent forms (3.15), (3.16), (3.20),
(6.7)-(6.9),(6.12), and (6.15) satisfies}
$$\int_{-\infty}^\infty dr\,[u_r(r,t)]^2=16\,\text{Tr}[A],\tag 7.16$$
{\it and $u(x,t)$ converges to an integer
multiple of $(2\pi)$ as $x\to-\infty.$}

\noindent PROOF: From (7.6) and
the second equation
in (7.3) we see that
$$\int_x^\infty  dr\,[u_r(r,t)]^2=8
CE(x,t)^{-1}Pe^{-\beta}B,$$
and hence with the help of Proposition~6.1, (5.1), and (6.10) we get
$$\int_{-\infty}^\infty  dr\,[u_r(r,t)]^2=8CP^{-1}B=
8\text{Tr}[BCP^{-1}]=
8\text{Tr}[(AP+PA)P^{-1}]=16\text{Tr}[A],$$
yielding (7.16).
By taking the time derivative of both sides of
(7.16), we get
$$0=\int_{-\infty}^\infty dr\,
u_r(r,t)\,u_{rt}(r,t)=
\int_{-\infty}^\infty dr\,
u_r(r,t)\,\sin(u(r,t))=\cos(u(-\infty,t))-\cos(u(+\infty,t)),$$
which proves that $u(-\infty,t)$ is an integer multiple
of $(2\pi)$ because we use the convention that
$u(+\infty,t)=0.$ \qed

\vskip 10 pt
\noindent {\bf 8. TRANSMISSION COEFFICIENT AND NORMING CONSTANTS}
\vskip 3 pt

In this section we show that our exact solutions given in equivalent forms
(3.15), (3.16), (3.20), (6.7)-(6.9), (6.12), and (6.15)
correspond to zero reflection coefficients
in (1.4), we evaluate that corresponding Jost solution explicitly
in terms of our triplet
$(A,B,C),$ determine the transmission coefficient explicitly
in terms of the matrix $A,$
and we also relate our triplet to the norming constants for
(1.4) and to their time evolutions.
As we have seen in Section~4 there is no loss of generality in choosing our triplet in the
special form specified in Theorem~4.5, and hence in this section we will assume
that $(A,B,C)$ has the particular form given in (4.12)-(4.17).

The Jost solution $\psi(\lambda,x,t)$ satisfying the asymptotics (2.1)
is given, as in (2.9) of [6], by
$$\psi(\lambda,x,t)=\bmatrix 0\\
\noalign{\medskip}
e^{i\lambda x}\endbmatrix+\int_x^\infty dy\bmatrix K(x,y,t)\\
\noalign{\medskip}
G(x,y,t)\endbmatrix e^{i\lambda y},\tag 8.1$$
where $K(x,y,t)$ and $G(x,y,t)$ are the quantities in
(3.17) and (7.6), respectively. Using (3.17) and (7.6) in (8.1) we obtain
$$\psi(\lambda,x,t)=e^{i\lambda x}\bmatrix iCE(x,t)^{-1}(\lambda I+iA)^{-1}B\\
\noalign{\medskip}
1-iC E(x,t)^{-1}P e^{-\beta}(\lambda I+iA)^{-1}B\endbmatrix,\tag 8.2$$
where $E$ and $\beta$ are the quantities appearing in
(3.18) and (3.14), respectively. With the help of Propositions~5.1 and 5.2,
by taking the limit of (8.2) as $x\to-\infty$ and by comparing the result with
(2.2), we see that $L(\lambda,t)=0,$ and hence [1,4,32] also $R(\lambda,t)=0,$ and
$$\displaystyle\frac{1}{T(\lambda)}=1-iCP^{-1}(\lambda I+iA)^{-1}B.\tag 8.3$$
Using (5.1), with the help of Proposition~4.2 of [6], we can invert (8.3) to
get
$$T(\lambda)=1+iC(\lambda I-iA)^{-1}P^{-1}B.\tag 8.4$$
By using Proposition~4.3 of [6] and (5.1), we can write (8.4) as the ratio
of two determinants as
$$T(\lambda)=\displaystyle\frac{\det(\lambda I+iA)}{\det(\lambda I-iA)}.\tag 8.5$$

Having determined the transmission coefficient $T$ in terms of
the matrix $A$ appearing in (4.12), let us clarify the relationship
between $A$ and the poles and zeros of $T$ in $\bold C^+.$ From (8.5)
we see that the zeros and poles of
$T$ occur exactly at the eigenvalues of $(-iA)$ and
of $(iA),$ respectively, and
that the poles of $T$ occur either
on the positive imaginary axis or they are pairwise symmetrically
located
with respect to the imaginary axis
in $\bold C^+.$
A comparison of $T$ given in
(8.5) with $A_j$ given in (4.14) shows that a
bound-state pole $\lambda_j$ of
$T$ located on the positive imaginary axis is related
to the eigenvalue $\omega_j$ of
$A_j$ in the form $\lambda_j=i\omega_j.$
A comparison of the poles of
$T$ given in (8.5) with $A_j$ of (4.16) reveals
the relationship between the poles off
the imaginary axis and the real constants
$\alpha_j$ and $\beta_j$ appearing in $A_j;$
namely, the pair of bound-state poles of $T$
symmetrically located with respect to the imaginary axis
in $\bold C^+$ occur at $\lambda=\lambda_j$ and
$\lambda=-\lambda_j^*,$ where
$$\lambda_j=-\beta_j+i\alpha_j,\qquad
-\lambda_j^*=\beta_j+i\alpha_j.\tag 8.6$$

Having clarified the relationship between the matrix $A$
appearing in (4.12) and
the bound-state poles in
$\bold C^+$
of the transmission coefficient $T,$ let us now discuss
the relationship between the bound-state norming constants and
the row vector $C$ appearing in (4.12).
In case of nonsimple bound-state poles of $T,$ the bound state norming
constants can be introduced [6,12] in such a way that
the generalization from the simple to the nonsimple
bound states is the most natural. The summation term in (2.5)
assumes that there are $n$ simple bound-state poles of
$T$ at $\lambda=\lambda_j$ with the norming constants
$c_je^{-it/(2\lambda_j)}.$ Let us now generalize
it to the case where each bound-state pole
$\lambda_j$ has multiplicity $n_j,$ i.e. when
there are $n_j$ linearly independent solutions
to (1.4) for $\lambda=\lambda_j.$
The most natural generalization is obtained
by the association
$$c_j\mapsto C_j,\qquad \lambda_j\mapsto iA_j,
\qquad 1\mapsto B_j,$$
where $A_j,$ $B_j,$ $C_j$ are the matrices appearing in
(4.12). The summation term (2.5) then generalizes to one of
the equivalent terms given in the set of equalities
$$Ce^{-A^{-1}t/2}e^{-Ay}B=
\sum_{j=1}^n C_je^{-A_j^{-1}t/2}e^{-A_jy}B_j=
\sum_{j=1}^n \sum_{s=1}^{n_j}\displaystyle\frac{1}{(s-1)!}\,
y^{s-1}\theta_{js}(t)\,e^{i\lambda_j y},\tag 8.7$$
where $(A,B,C)$ is the special triplet appearing
in (4.12) and $\theta_{js}(t)$
are the norming constants associated with
the eigenvalue $\lambda_j$ with
multiplicity $n_j.$

 From (8.7) we observe the relationship
between the bound-state norming constants $\theta_{js}(t)$
and the vectors $C_j$ appearing in (4.12).
If $\lambda_j$ occurs on the positive imaginary axis, then
we see that $\theta_{js}(0)$ is the same as $c_{js}$ appearing
in (4.13) and hence the time evolution
$\theta_{js}(0)\mapsto \theta_{js}(t)$ is governed by
$$\bmatrix \theta_{jn_j}(t)&\cdots &\theta_{j1}(t)\endbmatrix =
\bmatrix \theta_{jn_j}(0)&\cdots &\theta_{j1}(0)\endbmatrix e^{-A_j^{-1}t/2},\tag 8.8$$
where $A_j$ is the matrix obtained as in (4.14) by using $\omega_j=-i\lambda_j$ there.
We note that the norming constants $\theta_{js}(t)$
are all real (positive, negative,
or zero) with the understanding that $c_{jn_j}(t)\ne 0.$

Because of the real valuedness
stated in (2.6), if
the bound-state pole $\lambda_j$ of $T$ occurring
off the positive
imaginary axis has
$\theta_{js}(t)$ as the norming constants, then the bound-state
pole occurring at $(-\lambda_j^*)$ has
$\theta_{js}(t)^*$ as the norming constants.
In this case (8.6) holds, and a comparison
of (8.8) with (4.15) and (4.16) reveals that
the contribution from the pair $\lambda_j$ and
$(-\lambda_j^*)$ is given by one of the equivalent
forms
$$\sum_{s=1}^{n_j}\displaystyle\frac{1}{(s-1)!}
\left[\theta_{js}(t)\,y^{s-1}e^{i\lambda_j y}
+\theta_{js}(t)^*\,y^{s-1}e^{-i\lambda_j^* y}\right]
=C_je^{-A_jy-A_j^{-1}t/2}B_j,$$
where $(A_j,B_j,C_j)$ is the real triplet of size $2n_j$
appearing in (4.15) and (4.16). Thus, we see that
the real constants $\epsilon_{js}$ and $\gamma_{js}$ appearing
in (4.15) are related to the real and imaginary
parts of the norming constants $\theta_{js}(t)$ as
$$\epsilon_{js}=\text{Re}[\theta_{js}(0)],
\qquad \gamma_{js}=-\text{Im}[\theta_{js}(0)].$$
Defining the real $1\times (2n_j)$ vector
$$\theta_j(t):=\bmatrix -\text{Im}[\theta_{jn_j}(t)]&\text{Re}[\theta_{jn_j}(t)]
&\cdots&\cdots&-\text{Im}[\theta_{j1}(t)]
&\text{Re}[\theta_{j1}(t)]\endbmatrix,$$
we obtain the time evolution
$\theta_{js}(0)\mapsto \theta_{js}(t)$ as
$$\theta_j(t)=\theta_j(0)\,
e^{-A_j^{-1}t/2},$$
where
$A_j$ is the $(2n_j)\times (2n_j)$ matrix appearing in (4.16).

Let us note that, by using (8.8), we can describe the time evolution of the (complex)
norming constants $\theta_{js}(t)$ for $s=1,\dots,n_j$
corresponding to the complex
$\lambda_j$ given in (8.6) by simply
replacing the real matrix $A_j$ of size $n_j\times n_j$ given in (4.14)
with a complex-valued $A_j$ of the same size.
That complex $A_j$ is simply obtained by replacing
$\omega_j$ in (4.14) by the complex quantity
$(-i\lambda_j).$ In that case, the time evolution of the
norming constants $\theta_{js}(t)^*$ for $s=1,\dots,n_j$
corresponding to the complex
$-\lambda_j^*$ given in (8.6) is simply obtained by taking the complex conjugate of both sides of (8.8).

In short, in the most general case the summation term in (2.5) is given by the expression
$Ce^{-Ay-A^{-1}t/2},$ where the triplet $(A,B,C)$ has the form
(4.12).

\vskip 10 pt
\noindent {\bf 9. EXAMPLES}
\vskip 3 pt

\noindent {\bf Example 9.1} The triplet $(A,B,C)$ with
$$A=\bmatrix a\endbmatrix,\qquad B=\bmatrix 1\endbmatrix,\qquad C=\bmatrix c\endbmatrix,$$
where $a>0$ and $c\ne 0,$ through the use of (3.19) and (6.3), yields
$$P=\bmatrix\displaystyle\frac{c}{2a}\endbmatrix,\qquad M=\bmatrix\displaystyle\frac{c}{2a}\, e^{-2ax-t/(2a)}\endbmatrix,$$
and hence from (6.7) we get
$$u(x,t)=-4\tan^{-1}\left(\displaystyle\frac{c}{2a}\, e^{-2ax-t/(2a)}\right).\tag 9.1$$
If $c>0,$ the solution in (9.1) is known as a ``kink" [25]; it moves
to the left with speed $1/(4a^2)$ and $u(x,t)\to -2\pi$ as $x\to-\infty.$
If $c<0,$ the solution in (9.1) is known as an ``antikink" [25]; it moves
to the left with speed $1/(4a^2)$ and $u(x,t)\to 2\pi$ as $x\to-\infty.$

\noindent {\bf Example 9.2} The triplet $(A,B,C)$ with
$$A=\bmatrix a&b\\
-b&a\endbmatrix,\qquad B=\bmatrix 0\\
1\endbmatrix,\qquad C=\bmatrix c_2&c_1\endbmatrix,$$
where $a>0,$ $b\ne 0,$ and $c_2\ne 0,$
through the use of (3.19), (6.3), and (6.9) yields
$$u(x,t)=-4\tan^{-1}\left(\displaystyle\frac{\text{num}}{\text{den}}\right),\tag 9.2$$
where
$$\text{num}:=8a^2 e^{a\zeta_+}
\left[(a c_1-b c_2)\cos(b\zeta_-)-(b c_1+a c_2)\sin(b\zeta_-)\right],$$
$$\text{den}:=b^2(c_1^2+c_2^2)+16 a^2(a^2+b^2) e^{2a\zeta_+},\qquad
\zeta_\pm:=2x\pm\displaystyle\frac{t}{2(a^2+b^2)}.$$
The solution in (9.2) corresponds to a ``breather" [25] and
$u(x,t)\to 0$ as $x\to-\infty.$ For example, the choice
$a=1,$ $b=2,$ $c_1=2,$ $c_2=1$ simplifies (9.2) to
$$u(x,t)=4\tan^{-1}\left( \displaystyle\frac{2 e^{2x+t/10}\sin(4x-t/5)}{1+4e^{4x+t/5}}\right).$$

\noindent {\bf Example 9.3} The triplet $(A,B,C)$ with
$$A=\bmatrix a_1&0\\
0&a_2\endbmatrix,\qquad B=\bmatrix 1\\
1\endbmatrix,\qquad C=\bmatrix c_1&c_2\endbmatrix,$$
where $a_1$ and $a_2$ are distinct positive constants,
and $c_1$ and $c_2$ are real nonzero constants, by proceeding the same way as in the previous
example, yields (9.2) with
$$\text{num}:=2(a_1+a_2)^2\left(
a_1 c_2 e^{2 a_1 x+t/(2 a_1)}
+a_2 c_1 e^{2 a_2 x+t/(2 a_2)}\right),$$
$$\text{den}:=-(a_1-a_2)^2 c_1 c_2+4 a_1 a_2 (a_1+a_2)^2
e^{(a_1+a_2)(2x+t/(2 a_1a_2))}
.\tag 9.3$$
If $(c_1 c_2)<0$ then the quantity
in (9.3) never becomes zero; the corresponding solution
is known as a ``soliton-antisoliton" [25] interaction.
On the other hand, if $(c_1 c_2)>0$ then
the quantity in (9.3) becomes zero on a curve
on the $xt$-plane and the corresponding solution
is known as a ``soliton-soliton" [25] interaction.
For example, the choice $a=1,$ $b=2,$ $c_1=\pm 1,$ $c_2=\mp 1$ yields
$$u(x,t)=\pm 4\tan^{-1}\left(\displaystyle\frac{18 e^{2x+t/2}-36 e^{4x+t/4})}
{1+72 e^{6x+3t/4}}
\right),$$
with $u(x,t)\to 0$ as $x\to-\infty.$
On the other hand, the choice $a=1,$ $b=2,$ $c_1=\pm 1,$ $c_2=\pm 1$ yields
the solution
$$u(x,t)=\mp 4\tan^{-1}\left(\displaystyle\frac{18e^{2x+t/2}+36 e^{4x+t/4})}
{-1+72 e^{6x+3t/4}}
\right),$$
with $u(x,t)\to \mp 4\pi$ as $x\to-\infty.$

\noindent {\bf Example 9.4} The triplet $(A,B,C)$ with
$$A=\bmatrix a&-1&0\\
0&a&-1\\
0&0&a\endbmatrix,\qquad B=\bmatrix 0\\
0\\
1\endbmatrix,\qquad C=\bmatrix c_3&c_2&c_1\endbmatrix,$$
where $a>0,$
and $c_1,$ $c_2,$ $c_3$ are
real constants with $c_3\ne 0,$ by proceeding the same way as in the previous
example, yields $u(x,t)$ in the form of (9.2),
where
$$\text{num}:=c_3^3 e^{-4a x-t/a}+32 g
,\qquad
\text{den}:=
4 a e^{-2ax-t/(2a)}[128 a^8 e^{4ax+t/a}+h_1+h_2],$$
$$g:=(8 a^4 c_1+8 a^3 c_2+8 a^2 c_3)-(4 a^2 c_2+8a c_3)t
+c_3 t^2+(16 a^4 c_2+16 a^3 c_3)x-8 a^2 c_3 xt+16 a^4 c_3 x^2,$$
$$h_1:=(8 a^4 c_2^2-8 a^4 c_1 c_3+16 a^3 c_2 c_3+14 a^2 c_3^2)-
(4a^2 c_2 c_3+4a c_3^2)t,$$
$$h_2:=c_3^2 t^2+(16 a^4 c_2 c_3+32 a^3 c_3^2)x-8 a^2 c_3^2 tx
+16 a^4 c_3^2 x^2.$$
The choice $a=1,$ $c_1=-1,$ $c_2=-1,$ $c_3=-2$ yields
$$u(x,t)=-4\tan^{-1}
\left(\displaystyle\frac
{e^{-4x-t}+8(16-10t+t^2+24x-8tx+16x^2)}
{2 e^{-2x-t/2}[32 e^{4x+t}+20-6t+t^2-8tx+40x+16x^2]}
\right),$$
with $u(x,t)\to 2\pi$ as $x\to-\infty.$
On the other hand, the
choice $a=1,$ $c_1=0,$ $c_2=0,$ $c_3=1$ yields
$$u(x,t)=-4\tan^{-1}
\left(\displaystyle\frac
{e^{-4x-t}+32(8-8t+t^2+16x-8tx+16x^2)}
{4 e^{-2x-t/2}[128 e^{4x+t}+14-4t+t^2-8tx+32x+16x^2]}
\right),$$
with $u(x,t)\to -2\pi$ as $x\to-\infty.$

\vskip 5 pt

\noindent{\bf Acknowledgments}.
One of the authors (T.A.) is greatly indebted to the University of Cagliari for its
hospitality during a recent visit. This material is based in part upon work supported
by the Texas Norman Hackerman Advanced
Research Program under Grant
no. 003656-0046-2007,
the University of Cagliari,
the Italian Ministry of Education and Research (MIUR) under PRIN grant no.
2006017542-003, INdAM,
and the Autonomous Region of Sardinia
under grant L.R.7/2007 ``Promozione della ricerca
scientifica e dell'innovazione tecnologica in Sardegna."

\vskip 5 pt

\noindent {\bf REFERENCES}

\item{[1]} M. J. Ablowitz and P. A. Clarkson, {\it Solitons, nonlinear
evolution equations and inverse scattering,} Cambridge Univ. Press, Cambridge,
1991.

\item{[2]} M. J. Ablowitz, D. J. Kaup, A. C. Newell, and H. Segur,
{\it Method for solving the sine-Gordon equation,}
Phys. Rev. Lett. {\bf 30} (1973), 1262--1264.

\item{[3]} M. J. Ablowitz, D. J. Kaup, A. C. Newell, and H. Segur,
{\it The inverse scattering transform-Fourier analysis for nonlinear problems,}
Stud. Appl. Math. {\bf 53} (1974), 249--315.

\item{[4]} M. J. Ablowitz and H. Segur, {\it Solitons and the inverse
scattering transform}, SIAM, Philadelphia, 1981.

\item{[5]} T. Aktosun, T. Busse, F. Demontis, and C. van der Mee,
{\it Symmetries for exact solutions to the nonlinear Schr\"o\-din\-ger
equation,} J. Phys. A {\bf 43} (2010), 025202, 14 pp.

\item{[6]} T. Aktosun, F. Demontis, and C. van der Mee,
{\it Exact solutions
to the focusing nonlinear Schr\"o\-din\-ger equation,} Inverse Problems
{\bf 23} (2007), 2171--2195.

\item{[7]} T. Aktosun and C. van der Mee, {\it Explicit solutions to the
Korteweg-de Vries equation on the half-line,} Inverse Problems {\bf 22} (2006), 2165--2174.

\item{[8]} H. Bart, I. Gohberg, and M. A. Kaashoek,
{\it Minimal factorization of matrix and operator functions,} Birk\-h\"au\-ser, Basel, 1979.

\item{[9]} E. D. Belokolos, {\it General formulae for solutions of initial
and boundary value problems for sine-Gordon equation,}
Theoret. Math. Phys. {\bf 103} (1995), 613--620.

\item{[10]} E. Bour, {\it Th\'eorie de la d\'eformation des surfaces,}
J. \'Ecole Imp\'eriale Polytech. {\bf 19} (1862), No. 39, 1--48.

\item{[11]} P. Bowcock, E. Corrigan, and C. Zambon, {\it Some
aspects of jump-defects in the quantum sine-Gordon model}, J. High Energy Phys.
{\bf 2005} (2005), 023, 35 pp. 

\item{[12]} T. N. Busse,
{\it Generalized inverse scattering transform for the
nonlinear Schr\"o\-din\-ger equation,} Ph.D. thesis,
University of Texas at Arlington, 2008.

\item{[13]} G. Costabile, R. D. Parmentier, B. Savo,
D. W. McLaughlin, and
A. C. Scott, {\it Exact solutions of the sine-Gordon equation describing
oscillations in a long (but finite) Josephson junction,} Appl. Phys.
Lett. {\bf 32} (1978), 587--589.

\item{[14]} F. Demontis, {\it Direct and inverse scattering of the matrix
Zakharov-Shabat system,} Ph.D. thesis, University of Cagliari, Italy, 2007.

\item{[15]} F. Demontis and C. van der Mee, {\it Explicit solutions of the
cubic matrix nonlinear Schr\"o\-din\-ger equation,}
Inverse Problems {\bf 24} (2008),
025020, 16 pp.

\item{[16]} L. P. Eisenhart, {\it A treatise on
the differential geometry of curves
and surfaces,} Dover Publ., New York, 1960.

\item{[17]} F. C. Frank and J. H. van der Merwe, {\it One-dimensional
dislocations. I. Static theory,}
Proc. Roy. Soc. London A {\bf 198} (1949), 205--216.

\item{[18]} G. Gaeta, C. Reiss, M. Peyrard, and T. Dauxois, {\it Simple
models of non-linear DNA dynamics,}
Riv. Nuovo Cimento {\bf 17} (1994), 1--48.

\item{[19]} R. N. Garifullin, L. A. Kalyakin, and M. A. Shamsutdinov,
{\it Auto-resonance excitation of a breather in weak ferromagnets,}
Comput. Math. Math. Phys. {\bf 47} (2007), 1158--1170.

\item{[20]}
I. Gohberg, S. Goldberg, and M. A. Kaashoek, {\it Classes of linear operators,}
Vol. I, Birkh\"auser, Basel, 1990.

\item{[21]} Chaohao Gu, Hesheng Hu, and Zixiang Zhou,
{\it Darboux transformations in
integrable systems,} Springer, Dordrecht, 2005.

\item{[22]} N. Jokela, E. Keski-Vakkuri, and J. Majumder, {\it Timelike
boundary sine-Gordon theory and two-component plasma,} Phys. Rev. D
{\bf 77} (2008), 023523, 6 pp.

\item{[23]} A. Kochend\"orfer and A. Seeger,
{\it Theorie der Versetzungen in
eindimensionalen Atomreihen. I.~Periodisch angeordnete Versetzungen,}
Z. Phys. {\bf 127} (1950), 533--550.

\item{[24]} V. A. Kozel and V. R. Kotliarov,
{\it Almost periodic solution of
the equation} $u_{tt}-u_{xx}+\sin u=0$,
Dokl. Akad. Nauk Ukrain. SSR Ser.~A,
{\bf 1976} (1976), 878--881. 

\item{[25]} G. L. Lamb, Jr., {\it Elements of soliton theory,}
Wiley, New York, 1980.

\item{[26]} E. Lennholm and M. H\"ornquist,
{\it Revisiting Salerno's sine-Gordon model of DNA: active regions and robustness,}
Phys. D {\bf 177} (2003), 233--241.

\item{[27]} K. M. Leung, D. W. Hone, D. L. Mills,
P. S. Riseborough, and S. E.
Trullinger, {\it Solitons in the linear-chain antiferromagnet,}
Phys. Rev. B {\bf 21} (1980), 4017--4026.

\item{[28]} P. Mansfield, {\it Solution of the initial value problem for the
sine-Gordon equation using a Kac-Moody algebra,}
Commun. Math. Phys. {\bf 98} (1985), 525--537.

\item{[29]} D. W. McLaughlin and A. C. Scott, {\it Perturbation analysis of
fluxon dynamics,} Phys. Rev. A {\bf 18} (1978), 1652--1680.

\item{[30]} H. J. Mikeska, {\it Solitons in a one-dimensional magnet with
an easy plane,} J. Phys. C {\bf 11} (1977), L29--L32.

\item{[31]} M. B. Mineev and V. V. Shmidt,
{\it Radiation from a vortex in a long
Josephson junction placed in an alternating electromagnetic field,}
Sov. Phys. JETP {\bf 52} (1980), 453--457.

\item{[32]} S. P. Novikov, S. V. Manakov, L. B. Pitaevskii,
and V. E. Zakharov,
{\it Theory of solitons. The inverse scattering method,}
Plenum Press, New York, 1984.

\item{[33]}
E. Olmedilla, {\it Multiple pole solutions of the nonlinear
Schr\"odinger equation,} Phys. D {\bf 25} (1987), 330--346.

\item{[34]} C. P\"oppe, {\it Construction of solutions of the sine-Gordon
equation by means of Fredholm determinants,}
Phys. D {\bf 9} (1983), 103--139.

\item{[35]} N. R. Quintero and P. G. Kevrekidis,
{\it Nonequivalence of phonon
modes in the sine-Gordon equation,}
Phys. Rev. E {\bf 64} (2001), 056608, 4 pp.

\item{[36]} M. Rice, A. R. Bishop, J. A. Krumhansl, and S. E. Trullinger,
{\it Weakly pinned Fr\"olich charge-density-wave condensates: A new, nonlinear,
current-carrying elementary excitation,}
Phys. Rev. Lett. {\bf 36} (1976), 432--435.

\item{[37]} C. Rogers and W. K. Schief, {\it B\"ack\-lund and Darboux
transformations,} Cambridge Univ. Press, Cambridge, 2002.

\item{[38]} M. Salerno,
{\it Discrete model for DNA-promoter dynamics,} Phys.
Rev. A {\bf 44} (1991), 5292--5297.

\item{[39]} C. Schiebold,
{\it Solutions of the sine-Gordon equations coming
in clusters,} Rev. Mat. Complut. {\bf 15} (2002), 265--325.

\item{[40]} A. Seeger, H. Donth, and A. Kochend\"orfer, {\it Theorie der
Versetzungen in eindimensionalen Atomreihen. III.~Versetzungen, Eigenbewegungen
und ihre Wechselwirkung,} Z. Phys. {\bf 134} (1953), 173--193.

\item{[41]} A. Seeger and A. Kochend\"orfer,
{\it Theorie der Versetzungen in
eindimensionalen Atomreihen. II.~Beliebig angeordnete und beschleunigte
Versetzungen,} Z. Phys. {\bf 130} (1951), 321--336.

\item{[42]} R. Steuerwald, {\it \"Uber ennepersche Fl\"achen und
B\"ack\-lund\-sche Transformation,} Abh. Bayer. Akad. Wiss. (M\"unchen) {\bf 40} (1936), 1--105.

\item{[43]} L. V. Yakushevich, {\it Nonlinear physics of DNA,} 2nd ed.,
Wiley, Chi\-ches\-ter, 2004.

\item{[44]} V. E. Zakharov and A. B. Shabat, {\it Exact theory of
two-dimensional self-focusing and one-dimensional self-modulation of waves in
nonlinear media,} Sov. Phys. JETP {\bf 34} (1972), 62--69.

\item{[45]} V. E. Zakharov, L. A. Takhtadzhyan, and L. D. Faddeev, {\it Complete
description of solutions of the ``sine-Gordon" equation,} Soviet Phys. Dokl.
{\bf 19} (1975), 824--826.

\end